\documentclass[sn-mathphys,Numbered]{sn-jnl}% Math and Physical Sciences Reference Style

\usepackage{graphicx}%
\usepackage{multirow}%
\usepackage{amsmath,amssymb,amsfonts}%
\usepackage{amsthm}%
\usepackage{mathrsfs}%
\usepackage[title]{appendix}%
\usepackage{xcolor}%
\usepackage{textcomp}%
\usepackage{manyfoot}%
\usepackage{booktabs}%
\usepackage{algorithm}%
\usepackage{algorithmicx}%
\usepackage{algpseudocode}%
\usepackage{listings}%
\usepackage{siunitx}
\usepackage{gensymb}
\usepackage{caption}
\usepackage{float}
\usepackage[utf8]{inputenc}
\usepackage[T1]{fontenc}
\usepackage{bm}
\usepackage[capitalise,noabbrev]{cleveref}
\begin{document}

\title[Article Title]{\textbf{Designed spin-texture-lattice to control anisotropic magnon transport in antiferromagnets}}

%%=============================================================%%
%% Prefix	-> \pfx{Dr}https://www.overleaf.com/project/6515c44fa9d0bf17ca4efa44

\author*[1]{\fnm{Peter} \sur{Meisenheimer}}\email{meisep@berkeley.com}
\equalcont{These authors contributed equally to this work.}

\author[2]{\fnm{Maya} \sur{Ramesh}}\email{mr862@cornell.edu}
\equalcont{These authors contributed equally to this work.}

\author[1,3]{\fnm{Sajid} \sur{Husain}}\email{shusain@lbl.gov}
\equalcont{These authors contributed equally to this work.}

\author[4]{\fnm{Isaac} \sur{Harris}}\email{iaharris@berkeley.edu}

\author[5]{\fnm{Hyeon Woo} \sur{Park}}\email{0130phw@kaist.ac.kr}

\author[6]{\fnm{Shiyu} \sur{Zhou}}\email{shiyu\_zhou@brown.edu}

\author[4]{\fnm{Hossein} \sur{Taghinejad}}\email{h.taghinejad@berkeley.edu}

\author[1,3]{\fnm{Hongrui} \sur{Zhang}}\email{hongruizhang@berkeley.edu}

\author[9,10]{\fnm{Lane} \sur{W. Martin}}\email{lwmartin@rice.edu}

\author[4]{\fnm{James} \sur{Analytis}}\email{analytis@berkeley.edu}

\author[8]{\fnm{Paul} \sur{Stevenson}}\email{p.stevenson@northeastern.edu}

\author[11, 12]{\fnm{Jorge} \sur{\'I\~niguez-Gonz\'alez}}\email{jorge.iniguez@list.lu}

\author[5]{\fnm{Se Kwon} \sur{Kim}}\email{sekwonkim@kaist.ac.kr}

\author[2,13,14]{\fnm{Darrell G.} \sur{Schlom}}\email{schlom@cornell.edu}

\author[7]{\fnm{Lucas} \sur{Caretta}}\email{lucas\_caretta@brown.edu}

\author[15]{\fnm{Zhi} \sur{Yao}}\email{jackie\_zhiyao@lbl.gov}

\author*[1,3,4,9]{\fnm{Ramamoorthy} \sur{Ramesh}}\email{rr73@rice.edu}

\affil*[1]{\orgdiv{Department of Materials Science and Engineering}, \orgname{University of California}, \orgaddress{\city{Berkeley}, \state{CA}, \country{USA}}}

\affil[2]{\orgdiv{Department of Materials Science and Engineering}, \orgname{Cornell University}, \orgaddress{\city{Ithaca}, \state{NY}, \country{USA}}}
    
\affil[3]{\orgdiv{ Materials Science Division}, \orgname{Lawrence Berkeley National Laboratory}, \orgaddress{\city{Berkeley}, \state{CA}, \country{USA}}}

\affil[4]{\orgdiv{Department of Physics}, \orgname{University of California}, \orgaddress{\city{Berkeley}, \state{CA}, \country{USA}}}

\affil[5]{\orgdiv{Department of Physics}, \orgname{Korea Advanced Institute of Science and Technology (KAIST)}, \orgaddress{\city{Daejeon}, \country{South Korea}}}

\affil[6]{\orgdiv{Department of Physics}, \orgname{Brown University}, \orgaddress{\city{Providence}, \state{RI}, \country{USA}}}

\affil[7]{\orgdiv{School of Engineering}, \orgname{Brown University}, \orgaddress{\city{Providence}, \state{RI}, \country{USA}}}

\affil[8]{\orgdiv{Department of Physics}, \orgname{Northeastern University}, \orgaddress{\city{Boston}, \state{MA}, \country{USA}}}

\affil[9]{\orgdiv{Departments of Physics and Astronomy and Materials Science and NanoEngineering and Rice Advanced Materials Institute}, \orgname{Rice University}, \orgaddress{\city{Houston}, \state{TX}, \country{USA}}}

\affil[10]{\orgdiv{Department of Chemistry}, \orgname{Rice University}, \orgaddress{\city{Houston}, \state{TX}, \country{USA}}}

\affil[11]{\orgdiv{Materials Research and Technology Department}, \orgname{Luxembourg Institute of Science and Technology (LIST)}, \orgaddress{\city{Esch-sur-Alzette}, \country{Luxembourg}}}

\affil[12]{\orgdiv{Department of Physics and Materials Science}, \orgname{University of Luxembourg}, \orgaddress{\city{Belvaux}, \country{Luxembourg}}}

\affil[13]{\orgdiv{Kavli Institue for Nanoscale Science}, \orgname{Cornell University}, \orgaddress{\city{Ithaca}, \state{NY} \country{USA}}}

\affil[14]{\orgdiv{Leibniz-Institut f\"ur Kristallz\"uchtung}, \orgaddress{\city{Berlin}, \country{Germany}}}

\affil[15]{\orgdiv{Applied Mathematics and Computational Research Division}, \orgname{Lawrence Berkeley National Laboratory}, \orgaddress{\city{Berkeley}, \state{CA}, \country{USA}}}

%%==================================%%

\abstract{\textbf{Spin waves in magnetic materials are promising information carriers for future computing technologies due to their ultra-low energy dissipation and long coherence length. Antiferromagnets are strong candidate materials due, in part, to their stability to external fields and larger group velocities. Multiferroic aniferromagnets, such as BiFeO$_3$ (BFO), have an additional degree of freedom stemming from magnetoelectric coupling, allowing for control of the magnetic structure, and thus spin waves, with electric field. Unfortunately, spin-wave propagation in BFO is not well understood due to the complexity of the magnetic structure. In this work, we explore long-range spin transport within an epitaxially engineered, electrically tunable, one-dimensional (1D) magnonic crystal. We discover a striking anisotropy in the spin transport parallel and perpendicular to the 1D crystal axis. Multiscale theory and simulation suggests that this preferential magnon conduction emerges from a combination of a population imbalance in its dispersion, as well as anisotropic structural scattering. This work provides a pathway to electrically-reconfigurable magnonic crystals in antiferromagnets.}}


\maketitle
\renewcommand{\figurename}{Figure}

\section*{Introduction}\label{sec1}
In the push to design next-generation memory and logic technologies that utilize spin, antiferromagnets are particularly attractive due to their stability against external magnetic fields and large magnon propagation group velocities \cite{jungwirth_antiferromagnetic_2016, han_coherent_2023, jungwirth_multiple_2018}. One pathway to control magnon transport is by lithographically patterning periodic structures to create a so-called magnonic crystal \cite{qiu_magnon_2014, chumak_magnonic_2017, deng_magnon_2002, zivieri_collective_2011, chumak_spin-wave_2009, chumak_all-linear_2010}. A disadvantage of these extrinsic structures, however, is that the periodic structure is fixed. Their utility would be enhanced if they were tunable or reconfigurable \cite{wessler_dipolar_2022}. To this end, the interaction of spin waves with magnetic domain walls has been studied in ferromagnets \cite{hertel_domain-wall_2004, li_puzzling_2023, pirro_advances_2021, hamalainen_control_2018}, since domain walls can be written and moved using current-induced effects \cite{caretta_relativistic_2020, avci_interface-driven_2019, thomas_oscillatory_2006}. In order to design and create a magnonic crystal, however, periodic arrays of domain walls would need to be created. Here, we investigate magnon-transport behavior in a room temperature magnetoelectric with an intrinsically periodic, one-dimensional-cycloidal magnetic structure. Because this spin-texture-lattice is intrinsic to a multiferroic material and governed by the directionality of the polarization, we demonstrate that electric fields can be used to manipulate the magnonic crystal and control transport behavior. \\

For this study,  BiFeO$_\textrm{3}$ (BFO) is used as a model system. BFO is a well-studied antiferromagnetic oxide due to the strong coupling between polarization (ferroelectricity) and magnetic (antiferromagnetism) order at room temperature, enabling all-electric-field control of magnetization \cite{bibes_multiferroics_2008, heron_deterministic_2014, spaldin_advances_2019, rovillain_electric-field_2010}. Due to this interaction, BFO has garnered interest in the field of spintronics, especially with recent observations showing that electric-field control of spin transport is possible. \cite{parsonnet_nonvolatile_2022}. The fundamental magnetic order in BFO consists of G-type antiferromagnetism, which is modulated by broken symmetries stemming from the polarization and from antiferrodistortive FeO$_6$ octahedral rotations These phenomena give rise to a long-period ($\lambda \approx$ 65 nm) spin cycloid \cite{burns_experimentalists_2020}. In its equilibrium structure, this antiferromagnetic cycloid propagates along three symmetry-equivalent directions: $\bm{k}_1$ ($[\bar110]$), $\bm{k}_2$ ($[\bar101]$), and $\bm{k}_3$ ($[0\bar11]$), when polarization, $\bm{P}$, is parallel to $[111]$ \cite{zhong_quantitative_2022}, with Fe atomic moments rotating in the $(11\bar2)$, $(1\bar21)$, and $(\bar211)$, for $\bm{k}_1$, $\bm{k}_2$, $\bm{k}_3$, respectively \cite{sosnowska_spiral_1982}. The antiferromagnetic cycloid is then further perturbed by the Dyzaloshinskii-Moriya interaction (DMI), canting the moments slightly out of the $\bm{P}-\bm{k}$ plane and resulting in a spin-density wave ($\bm{M}$) with the same period as the antiferromagnetic cycloid \cite{ramazanoglu_local_2011, rahmedov_magnetic_2012}. This intrinsic, periodic magnetic structure in BFO can be modulated by an electric field, a capability attributed to its strong magnetoelectric coupling \cite{lebeugle_electric-field-induced_2008, gross_real-space_2017}. Despite its fundamental nature, the behavior of spin propagation within such a complex spin structure remains not fully understood. This gap in knowledge suggests that the cycloid's periodicity could have significant, yet unexplored, effects on magnon dispersion.
\\

The intimate coupling between magnetic order and ferroelectric polarization in BFO allows for the use of its ferroelectric structure as a tool to design periodic magnetic lattices, facilitating spin-wave transport. Previous studies have shown that the ferroelectric domain structure of thin-film BFO can be controlled using electrostatics and the epitaxial strain arising from the lattice mismatch to the substrate \cite{chu_nanoscale_2006}, leading to quasi-ordered arrays of either 71$\degree$ or 109$\degree$ domains \cite{martin_nanoscale_2008}. This directly influences the magnetic structure \cite{sando_crafting_2013, haykal_antiferromagnetic_2020}. Our current study demonstrates that by engineering these phenomena through the boundary conditions imposed by electrostatics and lattice-mismatch strain, we can effectively guide the ferroelectric domain structure to modulate the spin cycloid. This manipulation results in the creation of a long-range, intrinsic, quasi-1D-ordered lattice. Such a lattice structure creates a gap in the magnon dispersion and introduces a significant spin-transport anisotropy, which is further tunable with an electric field.\\

\begin{figure}[h!]%
    \noindent\includegraphics[width=\textwidth]{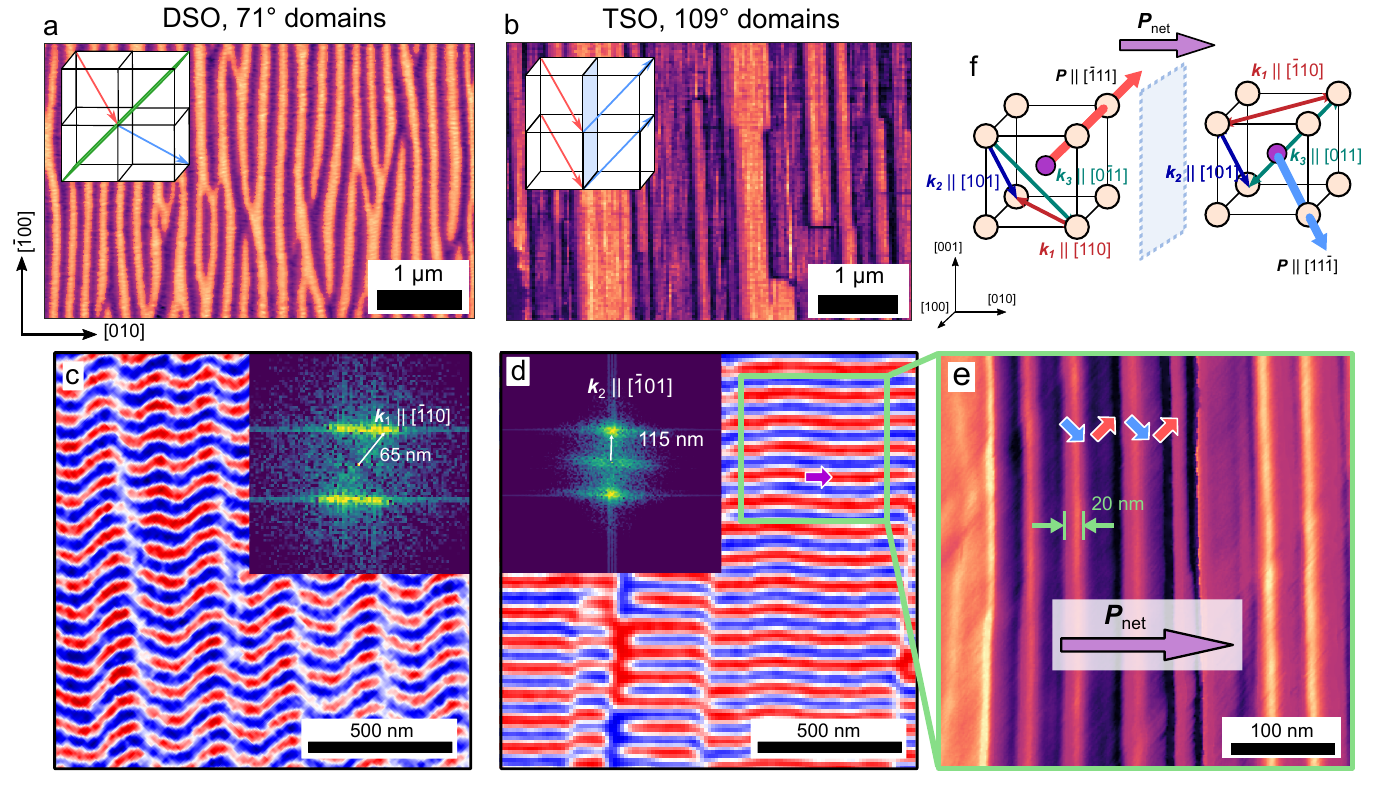}
    \caption{\textbf{1D ordered lattice.} In the two BFO variants, the ferroelectric domains are arranged into ground states containing 71$\degree$ (\textbf{a}) and 109$\degree$ (\textbf{b}) ferroelastic domain walls. Insets depict the two types of domain walls and their corresponding polarizations in the BFO unit cell. In films deposited on DSO substrates, the cycloid prefers the $\bm{k}_1$ variant, which results in a chevron-like pattern across the ferroelectric domains (\textbf{c}). For films on TSO, however, the domain structure allows selection of the other symmetry allowed axis, in this case $\bm{k}_2$ (\textbf{d}). Corresponding Fourier transforms are shown as insets revealing the period and symmetry of the magnetic structures. (\textbf{e}) Zoom of the area shown in green, where the cycloid is unperturbed by the fine, 10-30 nm ferroelastic domain walls. (\textbf{f}) This occurs due to the symmetry of $\bm{k}_2$ across the 109$\degree$ domain wall, where the directionality of the cycloid is preserved with the fixed net polarization $P_{net}$ imposed by the ferroelectric domains.}
    \label{Fig:struc}
\end{figure}

\section*{Designing the multiferroic structure}
To control the ferroelectric and magnetic structure of BFO, we synthesize epitaxial thin films on both DyScO$_3$ (DSO, $\sim$0.3\% compressive strain) and TbScO$_3$ (TSO, $\sim$0\% strain) substrates (Methods). From X-ray diffraction studies (Extended data \cref{supfig:xray}), films are confirmed to be $\sim$100 nm thick and of excellent crystalline quality. In addition to slightly different epitaxial strain states from the choice of substrate, these systems are chosen to engineer films with different ferroelectric domain structures. These films host characteristic stripe-like 71$\degree$ ferroelastic domains on DSO and 109$\degree$ ferroelastic domains on TSO \cite{chu_domain_2007}. Piezoresponse force microscopy (PFM) images (\cref{Fig:struc}a,b) demonstrate the two-variant stripe patterns along $[100]$. The nanoscale magnetic structures of these different strain and domain configurations are then investigated in real space using scanning nitrogen-vacancy (NV) magnetometry (Methods) \cite{gross_real-space_2017, haykal_antiferromagnetic_2020, zhong_quantitative_2022, meisenheimer_persistent_nodate}.\\

In films deposited on DSO (\cref{Fig:struc}c), NV magnetometry shows a chevron-like magnetic texture of the stray magnetic field, revealing a $\bm{k}_1$-oriented cycloid that switches direction with the change of the $\bm{P}$ directions across the 71$\degree$ ferroelastic domain walls. This is consistent with previous reports \cite{gross_real-space_2017} and is a result of the magnetoelastic anisotropy from the substrate lattice mismatch \cite{sando_crafting_2013, meisenheimer_persistent_nodate}. Strikingly, in the sample deposited on TSO (\cref{Fig:struc}d), only a single $\bm{k}$ variant of the cycloid is observed, projected along $[100]$. The longer period of these oscillations indicates that this corresponds to a $\bm{k}_2$, $[101]$ axis of the cycloid, which is now preferred due to the lower magnetoelastic energy from the smaller epitaxial strain of TSO. Importantly, this variant of the cycloid even persists across the approximately 20-30 nm wide 109$\degree$ ferroelectric domains and extends over long distances (up to at least 20 $\mu$m). This results in the formation of a quasi-long-range, 1D magnonic crystal, as illustrated in (\cref{Fig:struc}e). \\ 

This propagation of the $\bm{k}_2$ cycloid variant across ferroelectric domains can be understood through the symmetry of the 109$\degree$ ferroelastic domain walls. Starting with the 71$\degree$ domains in films on DSO substrates, the in-plane $\bm{k}_1$ directions are strongly magnetoelastically preferred. At the ferroelectric domain walls, where polarization P rotates from $[\bar111]$ to $[111]$, the cycloid wave vector directions must be shifted, for example, from $[110]$ to $[\bar110]$, to ensure the perpendicular relation $\bm{k}_1\perp \bm{P}$. This shift leads to a chevron-like pattern being favored as an equilibrium state. In contrast, in films on TSO substrates, the magnetoelastic energy is lowered, making the $\bm{k}_2$ and $\bm{k}_3$ variants closer in energy. Furthermore, considering the 109$\degree$ domain walls in the films on TSO substrates, assuming $\bm{k}_2$ is now an allowed direction, the propagation axis would not need to alter with the polarization change. This is because $[101]$ is perpendicular to both $[\bar111]$ and $[11\bar1]$ (\cref{Fig:struc}f), making $\bm{k}_2$ also a symmetry-favored equilibrium in this configuration. Consequently, $\bm{k_2}$ emerges as the globally preferred equilibrium state, leading to a single cycloid variant that persists across the 109$\degree$ ferroelastic domain walls, despite their $\sim$20-30 nm period \cite{chu_nanoscale_2006}.\\ 

\begin{figure}[h!]%
    \noindent\includegraphics[width=\textwidth]{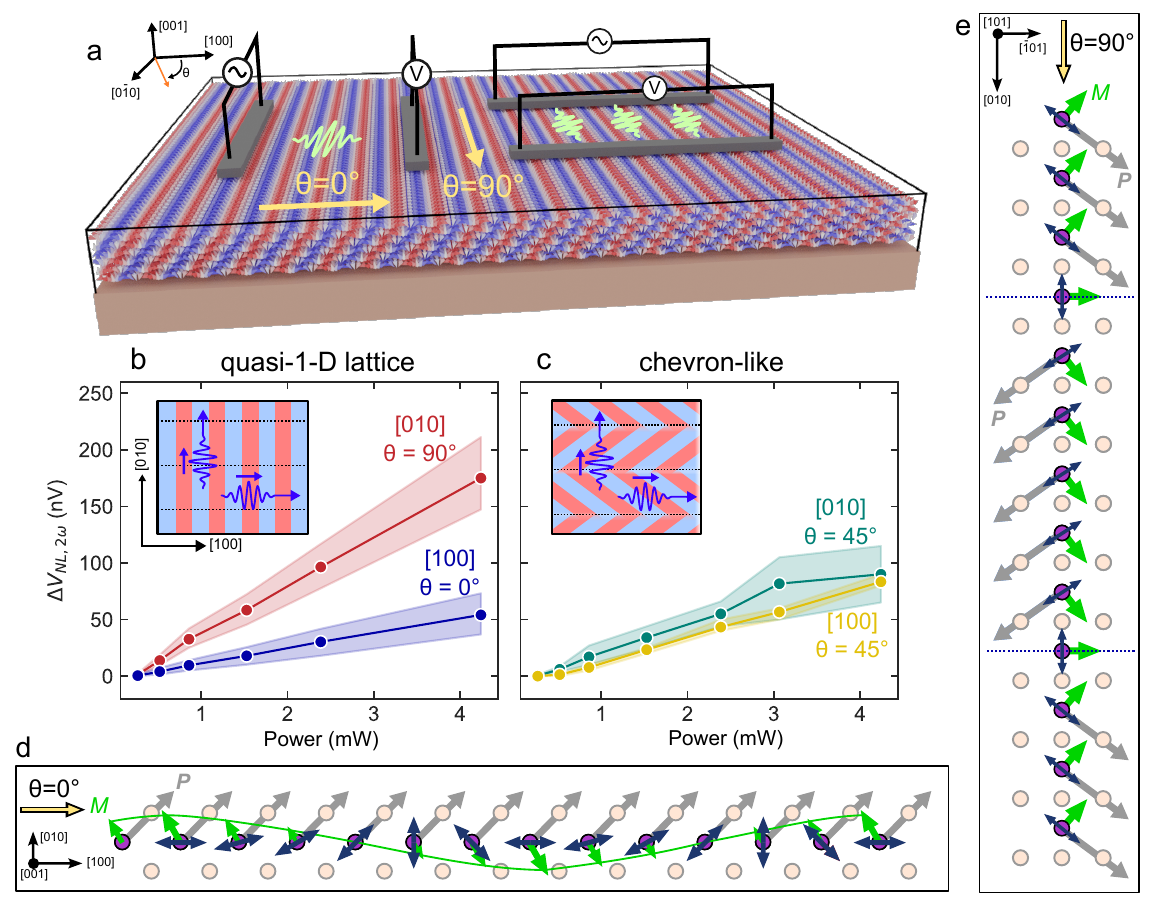}
    \caption{\textbf{Spin wave anisotropy.} (\textbf{a}) Schematic of the sample geometry of the test structures  $\parallel$ ($\theta =0\degree$) and $\perp$ ($\theta =90\degree$) to $\bm{k}$. These structures give rise to different magnon conductivities due to the directionality of the cycloid. In the quasi-1-D lattice of 109$\degree$ domains, (\textbf{b}), there is high conductivity along the ferromagnetically coupled direction, [010], and low conductivity along $\bm{k}$, [100]. Conversely, for the 71$\degree$ domain scenario, (\textbf{c}), the cycloid, and thus the conductivity, display no distinct anisotropy between the $\theta =0\degree$ and $90\degree$ directions. This reflects the magnetic cycloid's interaction with both the  109$\degree$ and 71$\degree$ ferroelastic domain walls. In the case of 109$\degree$ cycloid, viewed along the directions parallel to $[010]$ ($\theta=0\degree$), \textbf{d}), or $[100]$ ($\theta=90\degree$, \textbf{e}), the magnetic moments only deviate slightly across the ferroelectric domain walls, but cycle by $2\pi$ in the projected direction of $\bm{k}_2$. In \textbf{e}, the $109\degree$ ferroelastic domain walls are shown along with the deviation of $\bm{M}$ across the wall. In \textbf{e}, the antiferromagnetic Fe moments rotate in the $\bm{k}-\bm{P}$ plane, along with the magnitude of $\bm{M}$ traced by the green line. Unit cells are shown in an approximately 15:1 scale. The orientation of schematics \textbf{d} and \textbf{e} is viewed along the orthogonal directions shown in \textbf{a}.}
    \label{Fig:anis}
\end{figure}

\section*{Magnon transport}
To test spin transport with respect to the directions defined by the quasi-1D spin-spiral lattice, we fabricated test structures that are schematically shown in \cref{Fig:anis}a. These structures consist of source and detector wires aligned both parallel and perpendicular to $\bm{k}$. The magnon-induced voltage signal in these structures is compared along multiple directions with respect to $\bm{k}$ to show anisotropy, as well as in both the quasi-1D and chevron-like cycloid structures (\cref{Fig:struc}c). First, considering thermally excited magnons in a two-wire scheme, a current in the source wire is used to create a temperature gradient ($\nabla T$) in the device, causing thermally excited spin waves to propagate $\bm{q}\parallel -\nabla T$, where $\bm{q}$ is the magnon propagation direction, via the spin Seebeck effect \cite{parsonnet_nonvolatile_2022}. These magnons are then sensed in the detector wire through the inverse spin Hall effect (ISHE), which converts the spin current to a DC voltage, $\Delta V_{NL,2\omega}$ \cite{cornelissen_long-distance_2015, wesenberg_long-distance_2017}. To preserve the ferroelectric domains, we performed this measurement again by reversing the directionality of the thermal transport with respect to the in-plane polarization (details in Supp. Note 1). Confirmation that the signal is magnetic in origin comes from testing with different detector materials. For example, changing the detector wire from Pt to W (with a larger, opposite-signed spin-Hall angle compared to Pt) reverses the sign of $\Delta V_{NL,2\omega}$. Similarly, moving to a material with a much larger spin-Hall angle, such as SrIrO$_3$ (SIO) \cite{huang_novel_2021}, proportionally increases the ISHE voltage.\\

In films featuring the quasi-1D ordered cycloid lattice, a strong anisotropy is observed with respect to directions $\bm{q} \parallel \bm{k}$ and $\bm{q} \perp \bm{k}$. We define $\theta$ as the angle between $\bm{q}$ and $\bm{k}$. Notably, the ISHE voltage signal, indicative of spin wave conductivity, is approximately three times larger when $\theta=90\degree$, aligning with the cycloid's uniform direction. In contrast, samples with the alternating chevron-like structure, characterized by a net $\bm{k}_{net}=\sum \bm{k} =0$, exhibit minimal spin transport anisotropy (\cref{Fig:anis}c). Additionally, in the chevron-like samples, where locally $\theta = 45\degree$, $\Delta V_{NL,2\omega}$ lies between the values at $\theta = 90\degree$ and $0\degree$ seen in the 1D ordered system. This behavior can be initially understood by examining the nanoscopic magnetic spin structure within the cycloid as described schematically in \cref{Fig:anis}d,e. In a line cut $\perp \bm{k}$, the spins are modulated by the presence of the 109$\degree$ ferroelastic domain walls, but possess a persistent net magnetization component, illustrated in \cref{Fig:anis}e. In contrast, oriented parallel to $\bm{k}$, the spins rotate by a full $2\pi$ over the period of the cycloid (65 nm, \cref{Fig:anis}d). To further evaluate the physics associated with transport in the quasi-1D structure, we turn to both mesoscale phase-field simulations and model Hamiltonian calculations of spin transport $\parallel$ and $\perp \bm{k}$.\\

\begin{figure}[h!]%
    \noindent\includegraphics[width=\textwidth]{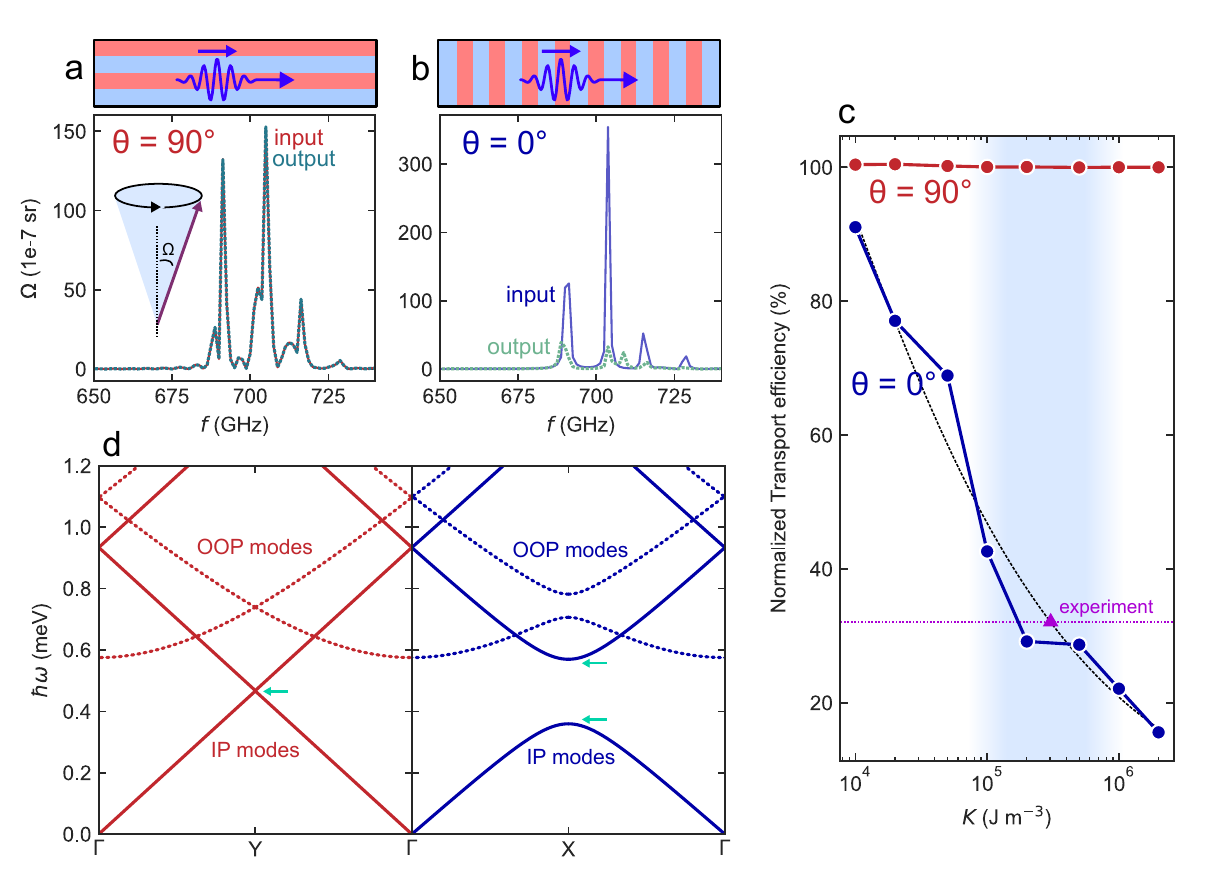}
    \caption{\textbf{Simulation of magnon conductivity}. Simulations of frequency-dependent magnon power when the spin wave is driven along the direction of $\bm{q}$, which is (\textbf{a}) perpendicular, and (\textbf{b}) parallel to, $\bm{k}$, respectively, in the quasi-1D lattice. Energy carried by the magnons is extracted as $\Omega\times f$, i.e., the product of the conical angle, $\Omega$, of the time-dependent procession of magnetization in Fourier space, and the frequency $f$. Input and output spectra are measured at the initialization location and far side of the simulated cell. The integral of this spectra is then reported as a relative conductivity, (\textbf{c}), simulated as a function of the 1D anisotropy, $K$, of the cycloid texture. The shaded region represents a range of energies reported in the literature, as well as values matched to the experimental and analytical results presented here. The horizontal dashed line corresponds to the measured efficiency, which is related to the plot by the intercept shown. (\textbf{d}) Calculated magnon dispersion, illustrating the emergence of a gap along the $\bm{q} \parallel \bm{k}$ direction which leads to the reduced spin wave conductivity. $X$ and $Y$ are the high symmetry points corresponding to the [100] ($\parallel \bm{k}$) and [010] ($\perp \bm{k}$), respectively.} 
    \label{Fig:theory}
\end{figure}

From phase-field simulations, we observe an anisotropy characterized by reduced spin-wave conductivity along the $\bm{q} \parallel \bm{k}$ direction, compared to $\bm{q} \perp \bm{k}$. Simulations are carried out by imposing a cycloidal magnetic ground state comparable to those observed experimentally and by initiating spin waves within a specified frequency range at one edge of the sample (Methods). The efficiency of these waves was quantified by comparing the magnetization precession at the point of origin and a defined distance from the source. In scenarios where $\bm{q} \perp \bm{k}$, the initial waveform exhibits negligible damping even at a distance equivalent to $10\lambda$ (\cref{Fig:theory}a). When $\bm{q} \parallel \bm{k}$, however, the spin wave exhibits significant attenuation over the same distance. This attenuation is evident in both the spectrum of the magnetization precession angle and its integrated energy across the spectrum, which is shown to be influenced by the magnitude of the anisotropy within the cycloid. As illustrated in \cref{Fig:theory}b, a scenario with an anisotropy of $K$ = 100 kJ m$^{-3}$ leads to approximately 50\% damping in spin-wave-energy magnitude over the same length scale. This transport efficiency can then be studied as a function of $K$ to compare with previous calculations and experimental results, defining a reasonable window for the value of $K$. The precession angles corresponding to other values of $K$ are shown in Extended data \cref{supfig:coneangle}.\\

%%% analytical

At the microscopic scale, analytical calculations of the momentum-dependent magnon dispersion in the quasi-1D lattice illustrate the difference in the magnon conduction $\perp \bm{k}$ and $\parallel \bm{k}$. Analytical solutions for the magnon bands of BFO are based on a continuum Hamiltonian whose density is given by~\cite{park_structure_2014}:
\begin{equation}
    \mathcal{H} = \mathcal{A}(\nabla \bm{\mathrm{n}})^2 - \alpha \bm{\mathrm{P}} \cdot [\bm{\mathrm{n}}(\nabla \cdot \bm{\mathrm{n}}) + \bm{\mathrm{n}} \times (\nabla \times \bm{\mathrm{n}})] - 2\beta M_0 \bm{\mathrm{P}} \cdot (\bm{\mathrm{m}} \times \bm{\mathrm{n}}) - K_u n^2_c + \lambda \bm{\mathrm{m}}^2 \, ,
\end{equation}
where the first term is the exchange energy, the second and third terms are DMI terms, the fourth term is the anisotropy, and the last term represents the suppression of the net magnetization due to strong antiferromagnetic coupling. The ground-state configuration can then be obtained (Supp. Note 3), showing the well-known anharmonic spin-cycloid structure \cite{park_structure_2014, sosnowska_origin_1995, kadomtseva_phase_2006}. Solving the time-dependent Euler-Lagrange equation for the small fluctuations, two spin-wave modes result: 1) An in-plane mode with spin fluctuations within the spin-cycloid plane and 2) an out-of-plane mode whose spin fluctuations are orthogonal to the spin-cycloid plane. By considering the quasi-2D geometry of the plane-wave-like spin texture of BFO and neglecting the Hamiltonian term $\propto \beta$ that is known to be much weaker than the rest of the terms in the Hamiltonian~\cite{park_observation_2014}, the band structures of in-plane (IP) magnons and out-of-plane (OOP) magnon modes are obtained analytically (\cref{Fig:theory}d). This clearly shows the formation of a gap in the dispersion along the $\bm{q}\parallel \bm{k}$ direction, due to the periodicity of the magnetic lattice. This gap indicates a lower density of states along this direction, resulting in an anisotropic, lower magnon conductivity.\\

%%% 1w

\begin{figure}[h!]%
    \noindent\includegraphics[width=\textwidth]{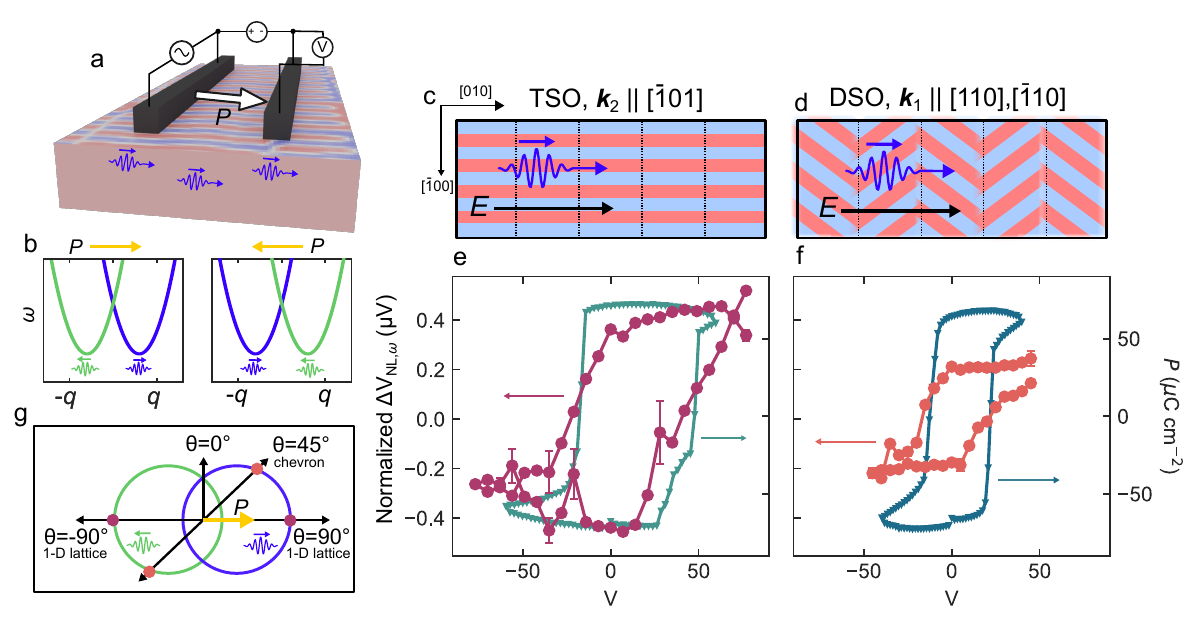}
    \caption{\textbf{Polarization control of magnon propagation.} (\textbf{a}) Schematic showing the non-local device geometry with magnons generated by injection of spin from the source wire. (\textbf{b}) We hypothesize that the non-reciprocity of the transport is due to a Rashba-like asymmetry of the magnon dispersion emerges from the crystal asymmetry, which is coupled directly to the ferroelectric polarization and DMI. (\textbf{c,d}) Illustration showing the sense of the cycloid with respect to the device geometry for films deposited on the two different substrates. In this orientation, the ferroelectric domain structure remains stationary with cycling electric field. (\textbf{e,f}) Switchable non-local signal overlaid with the ferroelectric hysteresis for both the single variant and chevron-like cases. (\textbf{g}) Combined with the above measurements, we hypothesize that the anisotropy can be explained by a split and anisotropic magnon dispersion, where the magnon conductivity maps to a surface with directions defined by $\bm{q}$ and $\bm{k}$. The distance from the origin is representative of the conductivity, and thus the population.}
    \label{Fig:efield}
\end{figure}

In parallel to measurements of thermally excited magnons, non-local transport measurements have been performed using injected spin currents via the spin-Hall effect (SHE) of the source wire \cite{cornelissen_long-distance_2015} (\cref{Fig:efield}a). In this case, the sign of the output signal will correspond directly to the sign of the input signal (i.e., a positive voltage will result in a positively polarized spin current, which then results in a positive voltage at the detector), rather than only the amplitude of the input (so-called first-harmonic and second-harmonic measurements). While the spin Hall angles of platinum and tungsten are too low to generate an appreciable ISHE signal at the 2 $\mu$m electrode spacing, the efficient spin-charge transduction of SIO allows detection of the SHE excited magnon signal \cite{huang_novel_2021}. In the device geometry where $\bm{q}$ is perpendicular to the domain walls, where electric fields allow the ferroelastic domain walls to remain largely stationary \cite{parsonnet_nonvolatile_2022, meisenheimer_persistent_nodate, heron_deterministic_2014} (Supp. \cref{supfig:domains}), combining the engineered spin texture with an epitaxial, large SHE oxide can further optimize functionality. This should allow for deterministic switching of the magnetic structure under a continuous rotation of $\bm{P}$ \cite{meisenheimer_persistent_nodate}.\\

In test structures using SIO electrodes, we observe a similar anisotropic spin transport, dependent on the underlying magnetic structure. Here, magnons are excited by the SHE and the ferroelectric polarization (noted $\bm{P}\uparrow$ and $\bm{P}\downarrow$, where the direction of magnon propagation, $\bm{q}\uparrow$, is fixed), is switched using an in-plane electric field. Upon switching the in-plane component of the polarization, where the direction of magnon propagation, $\bm{q}$, is fixed, the detected first-harmonic non-local voltage, $V_{NL,1\omega}$, switches hysteretically on the scale of $\sim$1 $\mu$V for a device spacing of 2 $\mu$m. The repeatable difference between the $\bm{P}\uparrow \bm{q}\uparrow$ and $\bm{P}\downarrow \bm{q}\uparrow$ states is indicative of a non-reciprocity in the magnon dispersion tied to the direction of $\bm{P}$. This can be qualitatively rationalized in the form of a Rashba-like, DMI-induced splitting of the spin-dependent magnon bands (sketched in \cref{Fig:efield}b) \cite{gitgeatpong_nonreciprocal_2017, dos_santos_nonreciprocity_2020, kawano_designing_2019, kawano_discovering_2019, shiomi_spin_2017}. The result is a $V_{NL,1\omega}$ that is both nonreciprocal and switchable.  Since the samples under investigation exhibit no net magnetic moment, the inference is that magnon non-reciprocity in BFO should stem from a chiral component of the magnetic cycloid, which interacts with $\bm{P}$. \\

Under an applied electric field, this hysteresis in $V_{NL,1\omega}$ follows the switching of $\bm{P}$, where $\bm{P}$ and $\Delta V_{NL,1\omega}$ both flip signs at the same coercive voltage (\cref{Fig:efield}d,e) \cite{bhattacharjee_ultrafast_2014}. As the ferroelastic domain walls remain stationary during this event (Supp. \cref{supfig:domains}),  a comparison can be made between samples with quasi-1D and chevron-like configurations (\cref{Fig:efield}c,d). In quasi-1D samples with a single $\bm{k}$ state, the magnitude of $V_{NL,1\omega} \perp \bm{k}$ is two-times higher than that of the samples showing a chevron-like magnetic cycloid. The magnitude of this scaling directly parallels the observations in \cref{Fig:anis}, where the chevron-like texture shows a two-times lower conductivity than in the single $\bm{k}$ samples when $\bm{q} \perp \bm{k}$. Taken together, these measurements allow us to sketch the approximate shape of the magnon dispersion projected into the plane of the film (\cref{Fig:efield}g), where $\theta=90\degree$ ($\bm{P}\perp \bm{k}$) is high conductivity, $\theta=0\degree$ ($\bm{P} \parallel \bm{k}$) is low, $\theta=45\degree$ is intermediate, and $\theta=90\degree$ and $\theta=-90\degree$ carry an opposite sign angular momentum.\\

These results demonstrate control of the magnetic structure of BFO and a method for effectively decoupling the magnetic structure from the ferroelectric domains. This has significant implications for magnonic devices using BFO, in that fundamental anisotropies of the cycloid structure can be used to optimize transport. The formation of a quasi-1D ordered magnonic lattice comprised of the spin cycloid should be of fundamental interest in terms of phase transitions in a 1D lattice and the possibility of obtaining long-range order in them. In this sense, future studies could be directed towards controlling the ferroelectric/ferroelastic domain structure through the use of vicinally cut substrates \cite{chu_nanoscale_2006}. 

\section*{Conclusion}
Using heteroepitaxy as a building block and imposing a combination of electrostatic and elastic constraints, we can strategically design both the ferroelectric and magnetic domain structure of multiferroic BFO. By doing so, a quasi-1D crystal of the antiferromagnetic spin cycloid can be chosen as the ground state. This anisotropic structure results in preferential spin-wave conduction orthogonal to its axis, due to the opening of a gap in the magnon dispersion created by the periodicity of the magnetic lattice. Leveraging magnetoelectric coupling, this magnonic crystal can be dynamically tuned using an electric field, allowing us to empirically map the magnon dispersion in the film plane with respect to the direction of ferroelectric polarization. This opens the door for the design of magnonic devices utilizing the anisotropy of the spin texture, as well as providing fundamental insights into the conduction of spin waves in complex magnetic systems. \\

\section*{Methods}
\textbf{Sample Fabrication} 100 nm thick BFO samples were deposited on DSO and TSO substrates using both molecular-beam epitaxy (MBE) and pulsed laser deposition (PLD). MBE films were grown by reactive MBE in a VEECO GEN10 system using a mixture of 80 \%  ozone (distilled) and 20 \% oxygen. Elemental sources of bismuth and iron were used at fluxes of \SI{1.5e14}{} and  \SI{2e13}{atoms \per\centi\meter\squared\second} respectively. All films were grown at a substrate temperature of \SI{675}{\celsius} and chamber pressure of 5 x 10$^{-6}$ Torr. PLD samples were deposited using a 248 nm KrF laser with $\sim$1.5 J cm$^{-2}$ at 700 $\degree$C and 90 mTorr O$_2$. Pt (15 nm) and W (15 nm) layers were deposited using magnetron sputtering at room temperature. SIO layers were deposited using MBE using a mixture of 80 $\%$ ozone and 20 $\%$ O$_2$ along with elemental sources of strontium and iridium. Due to the high temperatures needed to generate an iridium flux, an electron beam was used to evaporate iridium. The samples were grown at a substrate temperature of 695 $\degree$C and a background pressure of 1 x 10$^{-6}$ Torr. Devices were fabricated using standard optical lithography followed by ion milling.\\

\noindent\textbf{Sample Characterization} X-ray diffraction was performed on a PANalytical Xpert3 diffractometer. Piezoresponse force microscopy was performed on an Asylum MFP3D. NV magnetometry was performed on a Qnami ProteusQ system with a parabolic Quantilever MX+ diamond cantilever.\\

\noindent\textbf{Phase Field Simulations}
We employed MagneX, a substantial GPU-enabled micromagnetic simulation code package (available on \href{https://github.com/AMReX-Microelectronics/MagneX.git}{\sffamily \color{blue} \underline{GitHub}}), to phenomenologically explore the directionality of magnon transport in the cycloid. The time-domain Landau-Lifshitz-Gilbert equation is solved, incorporating exchange coupling, DMI coupling, anisotropy coupling, and demagnetization coupling.  A set of abstract ferromagnetic properties is utilized to phenomenologically represent the antiferromagnetic material.  The magnetization is pinned by an external applied magnetic field that mimics the cycloid texture. The Gilbert damping constant is set to be 0.0005, the exchange coupling constant is $3.0\times 10^{-12}$ J m$^{-1}$, and the DMI coupling constant is $1.0935\times 10^{-3}$ J m$^{-2}$. The anisotropic hard axis is set to the [111] direction, while the anisotropy constant is swept in the range of $1.0\times 10^{4}$ J m$^{-3}$  and $2.0\times 10^{6}$ J m$^{-3}$. The total magnetization is normalized to be one. On one side of the sample, a modified Gaussian pulse in the effective magnetic field is applied in both the $x$ and $y$ direction, mimicking injected spins from the top SHE electrode. The pulse is in the mathematical form of $e^{-(t-t_0)^2/2T_0^2}\cos(2\pi f_0 t)$, where $f_0$ is 700 GHz. The output port is on the opposite side of the input port in the in-plane direction. The simulated time-varying magnetization components are converted into the frequency domain with the Fourier transform, from which we calculate the cone angle of the magnetization. For more details, see Supp. Note 2. 
\\

\noindent\textbf{Analytical Calculations} See Supp. Note 3.\\

\noindent\textbf{Transport Measurements} Transport measurements were performed using 3-terminal devices where two were used to inject current and two for output. Two cross-terminals were used to apply electric field for polarization control. All operations were controlled through the in-house developed code. $V_{NL}$ was measured using an SR830 lock-in amplifier and locking to the second (in the case of thermally excited magnons) or first (in the case of spin Hall excited magnons) harmonic of the excitation voltage. An AC signal at 7 Hz was used for both first- and second-harmonic measurements. In the first-harmonic measurement, the magnitude of the injected current is chosen to be small to avoid the thermal drift in the output voltage.\\

\section*{Author Contributions}
PM designed experiments. MR and SH deposited samples. PM and SH performed PFM. PM, SZ, and PS performed NV microscopy. ZY performed phase field simulations. HWP and SKK performed analytical calculations. IH and HT fabricated devices. SH and IH performed spin transport. PM, MR, and SH wrote the manuscript. All authors have contributed to manuscript edits.\\

This work was primarily supported by the U.S. Department of Energy, Office of Science, Office of Basic Energy Sciences, Materials Sciences and Engineering Division under Contract No. DE-AC02-05-CH11231 (Codesign of Ultra-Low-Voltage Beyond CMOS Microelectronics (MicroelecLBLRamesh)) for the development of materials for low-power microelectronics. P.M., R.R., M.R. and D.G.S. additionally acknowledge funding from the Army Research Office under the ETHOS MURI via cooperative agreement W911NF-21-2-0162. S.Z. and L.C. acknowledge funding from Brown School of Engineering and Office of the Provost. P.S. acknowledges support from the Massachusetts Technology Collaborative, Award number 22032. Computations in this paper were performed using resources of the National Energy Research Scientific Computing Center (NERSC), a U.S. Department of Energy Office of Science User Facility operated under contract no. DE-AC02-05CH11231. The work performed at the Molecular Foundry was supported by the Office of Science, Office of Basic Energy Sciences, of the U.S. Department of Energy under Contract No. DEAC02-05CH11231. S.K.K. and H.W.P. acknowledge support from the Brain Pool Plus Program through the National Research Foundation of Korea funded by the Ministry of Science and ICT (2020H1D3A2A03099291). \\

The authors declare no competing interests\\  

Data relevant to this manuscript have been uploaded to zenodo under accession code XYZ. Other data is available from the corresponding author upon reasonable request.\\

% \bibliography{references}% common bib file

\newpage
\section*{Extended Data}
Extended data for "Designed spin-texture-lattice to control anisotropic magnon transport in antiferromagnets"

\setcounter{figure}{0}
\renewcommand{\figurename}{Extended Data Figure}

\begin{figure}[h!]%
    \noindent\includegraphics[width=\textwidth]{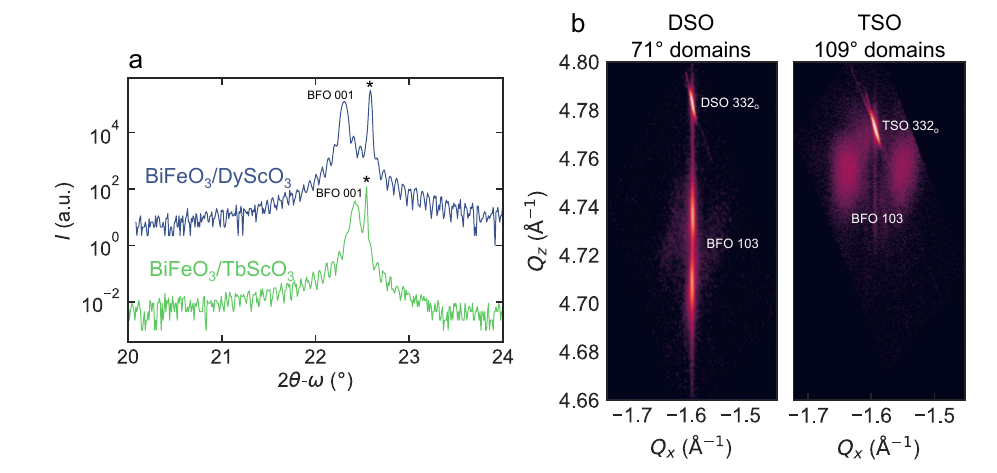}
    \caption{\textbf{X-ray diffraction a} $2\theta - \omega$ line scans of the two samples with Pendell\"osung fringes around the film peaks, showing the crystalline quality. \textbf{b} Reciprocal space maps of the $332_O$ diffraction peak. Differences are due to the different configrations of the rhombohedral BFO unit cell in the different domain configurations \cite{chu_nanoscale_2006}.}
    \label{supfig:xray}
\end{figure}

\begin{figure}[h!]%
    \noindent\includegraphics[width=\textwidth]{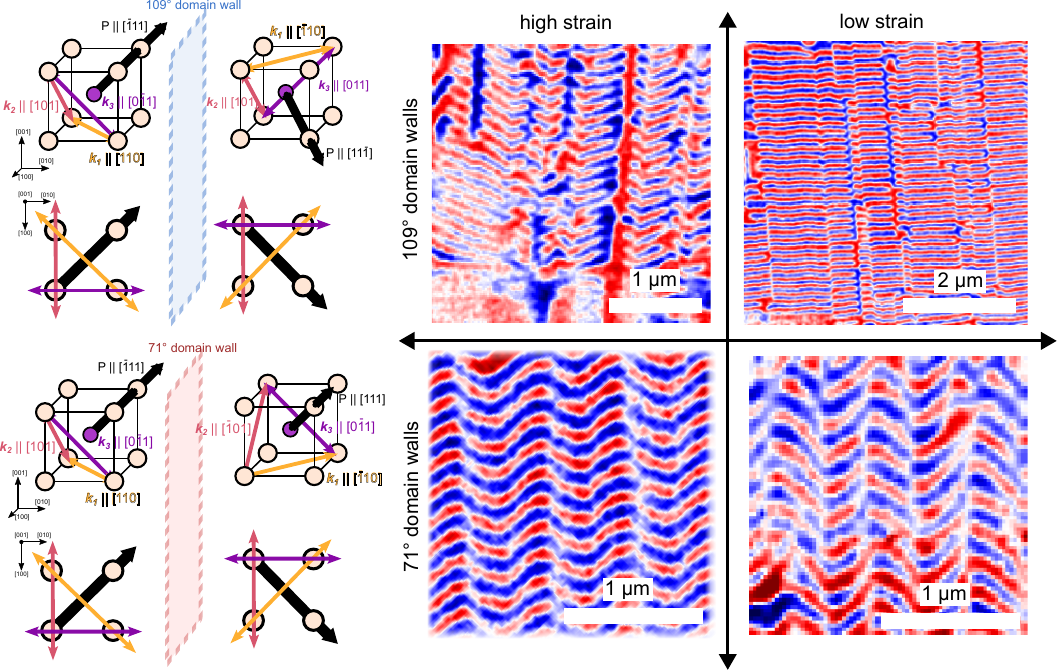}
    \caption{\textbf{Strain and domain wall competition a,b} Schematics showing the symmetry allowed $\bm{k}$ directions at 109$\degree$ and 71$\degree$ domain walls. \textbf{c} NV images showing the magnetic structure at different permutations of strain and domain wall symmetry. When strain is high, as in DSO, $\bm{k}_1$ is preferred, but becomes frustrated in 109$\degree$ domain walls, leading to competition between the type-i and type-ii cycloids. In samples with 71$\degree$ domains, it appears that $\bm{k}_1$ is also preferentially chosen as the ground state. It is only through the combination of low strain and allowed symmetry where $\bm{k}_2$, and the single variant state, is preferred.}
    \label{supfig:domains}
\end{figure}

\begin{figure}[h!]%
    \noindent\includegraphics[width=\textwidth]{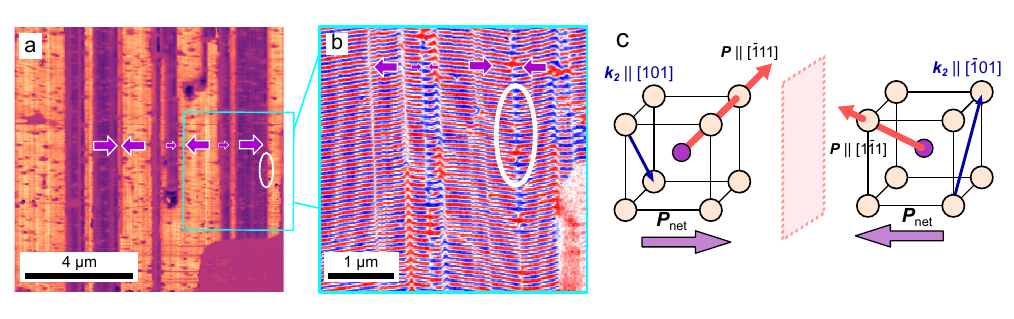}
    \caption{\textbf{Antiphase boundaries} The 109$\degree$ domain samples exhibit a hierarchical domain pattern characteristic of BFO.  Superimposed on the fine-scale 109$\degree$ domain ensemble with a spacing of 20-30nm, families of domains with net antiparallel in-plane polarization components are also observed. Comparing the ferroelectric polarization from PFM with the NV data, we observe discontinuities in the 1-D cycloid lattice that correlate with the family domain walls (\textbf{a,b}). Though $\bm{k}_2$ does not change across the fine-scale (20-30nm)  domain walls, it does change (e.g. $\bm{k}_2[101]$ to $\bm{k}_2'[\bar101]$) across these large domains with a different in-plane polarization (\textbf{c}). Though $\bm{k}_2$ and $\bm{k}_2'$ in these two domains are orthogonal to each other, the in-plane component of the cycloid does not change, giving the appearance of discontinuities (akin to an antiphase boundary in ordered alloys) when projected in-plane. When the material is electrically poled, these macro domain walls are annealed out of the sample.}
    \label{supfig:antiphase}
\end{figure}

\begin{figure}[h!]%
    \noindent\includegraphics[width=\textwidth]{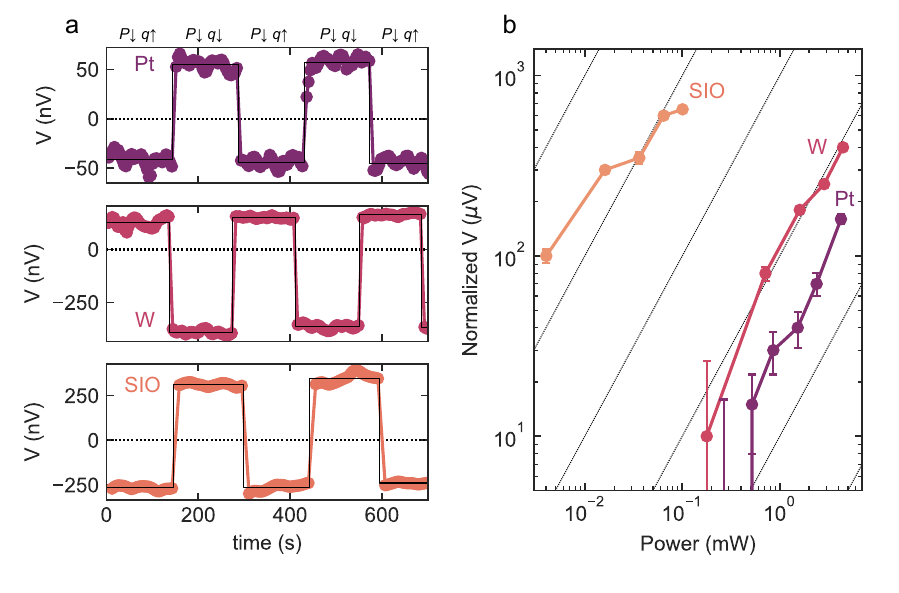}
    \caption{\textbf{Material and power dependence} \textbf{a} The thermally-driven non-local conductivity changes with the sign and magnitude of the spin Hall angle of the detector wire material (Pt, W, and SIO), demonstrating its magnetic origin. Symbols are the raw data and solid lines represent the time average. $P$ and $q$ are the polarization and the magnon-propagation direction, respectively. \textbf{b} The output voltage also scales with the magnitude of the spin Hall angle of the detector material. The dotted lines correspond to orders of magnitude.}
    \label{supfig:power}
    \centering
\end{figure}

\begin{figure}[h!]%
    \noindent\includegraphics[width=\textwidth]{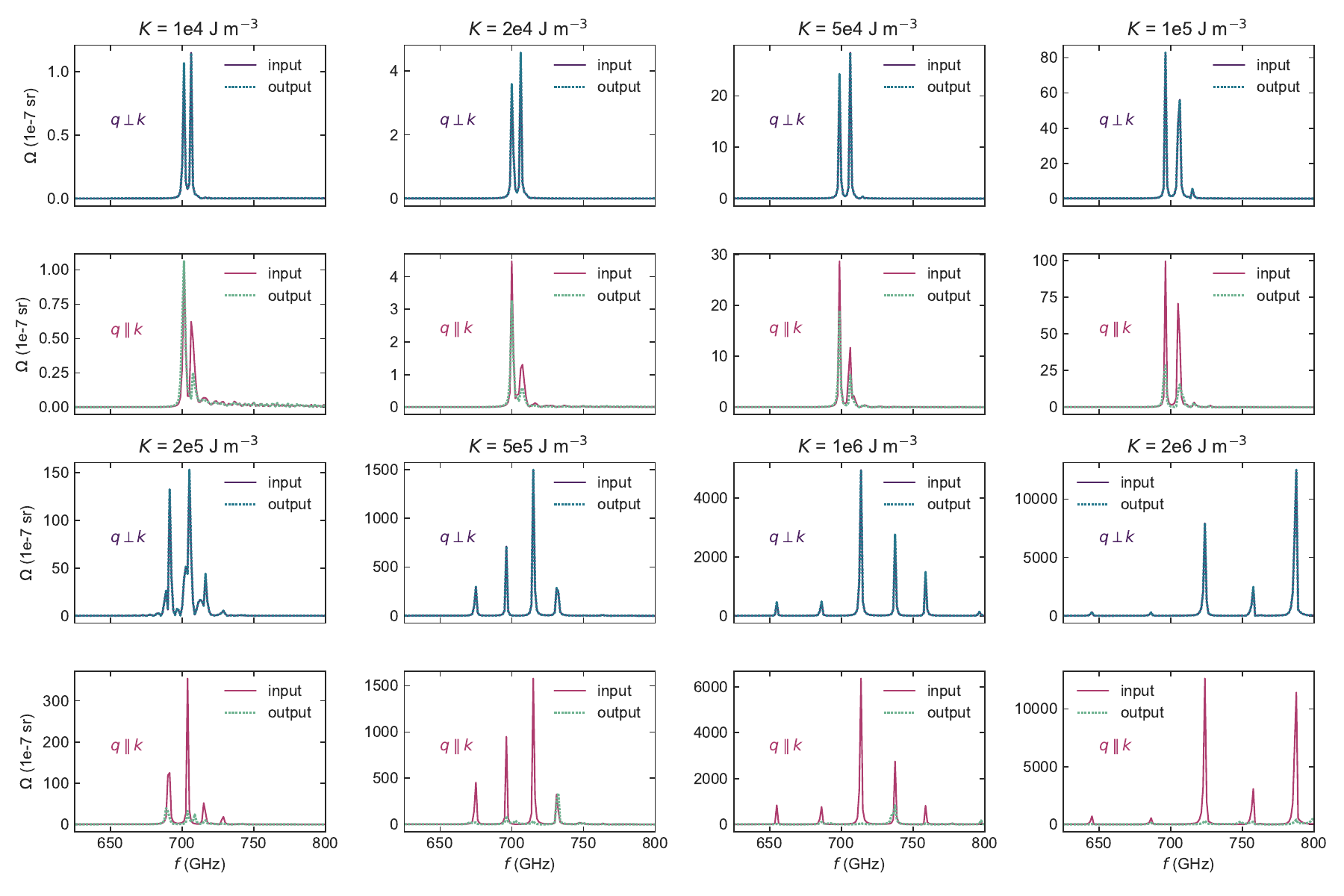}
    \caption{\textbf{Spectral dependence of magnetocrystalline anisotropy} Resonant spectra of excited spin waves extracted from phase field simulations in both the $\bm{q}\perp \bm{k}$ and $\bm{q}\parallel \bm{k}$ cases, used to calculate the efficiency scaling in  \cref{Fig:theory}c. Frequency dependent cone angles, $\Omega$, are calculated as the Fourier transforms of the time dependent magnetization procession. Anisotropy ($K$) values are intentionally chosen both above and below the expected range. }
    \label{supfig:coneangle}
    \centering
\end{figure}

\begin{figure}[h!]%
    \noindent\includegraphics[width=\textwidth]{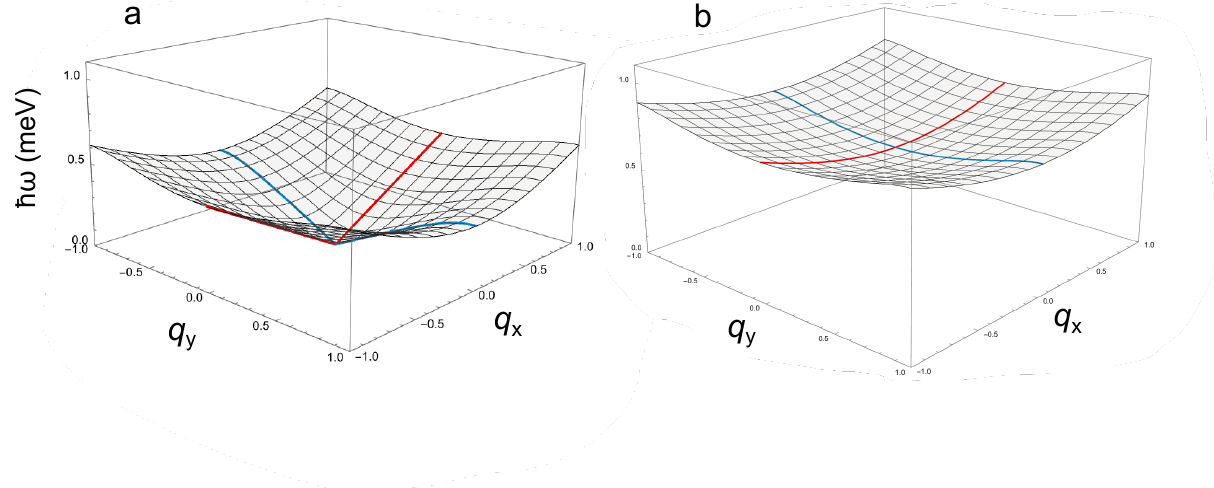}
    \caption{\textbf{2D magnon dispersion} visualizations of the of the lowest IP (\textbf{a}) and OOP (\textbf{b}) bands, showing both $\bm{q}_x \perp \bm{k}$ (red) and $\bm{q}_y \parallel \bm{k}$ (blue). The in-plane and out-of-plane mode corresponds to the spin
    fluctuations within the spin-cycloid plane and perpendicular to the spin-cycloid plane, respectively. The striking difference between these energy bands demonstrates the anisotropic magnon-dispersion within the spin cycloid 1D lattice.  }
    \label{supfig:disp}
    \centering
\end{figure}

\begin{figure}[h!]%
    \noindent\includegraphics[width=\textwidth]{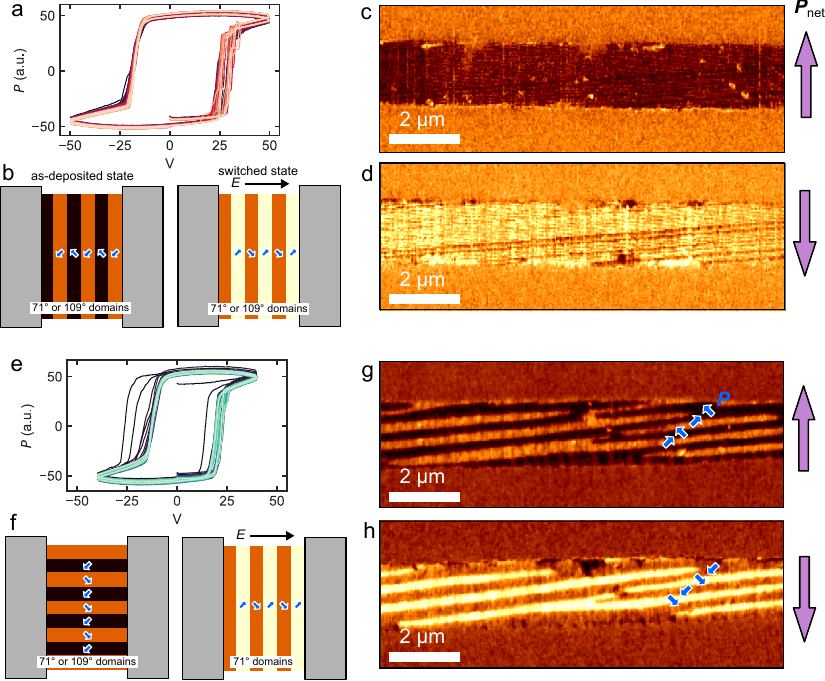}
    \caption{\textbf{Ferroelectric switching} Ferroelectric hysteresis loops (\textbf{a,e}) and corresponding PFM (\textbf{c,d,g,h}) of the poled ferroelectric states. In both cases, devices can be electrically poled with $E \perp$ the domain walls and conserve the domain structure (\textbf{b}). When $E \parallel$ to to the domain walls, (\textbf{h}) however, the domains reorient into the striped 71$\degree$ structure such that $E \perp$ the domain walls, regardless of the starting configuration (\textbf{f}). This can be observed in the hysteresis loop in d, where, starting from the as-deposited state, parallel 109$\degree$ domains are progressively transformed into perpendicular 71$\degree$ domains as the device is cycled 3-4 times.}
    \label{supfig:devices}
\end{figure}

\end{document}

\title[Article Title]{Supplementary Information for Designed spin texture lattice to control magnon transport in antiferromagnets}

\maketitle
\renewcommand{\figurename}{Supp. Figure} % fix caption
%%==================================%%

\section*{Supplementary Note 1: Magnon transport measurements}
In thermally excited magnon transport experiments, the magnitude of non-local inverse spin Hall voltage ($V_{NL}$) is reported as the difference ($\Delta V_{NL}$) between two measurement configurations where the propagation direction, $q$ from the thermal gradient, $\nabla T$, is changed and the ferroelectric polarization ($P$) remains fixed. This was chosen due to the response of domains to an electric field parallel to the domain wall, which leads to reconfiguration of the domain structure. The application of an electric field parallel to the two-variant domain walls transforms the domain structure into 71$\degree$ domains running perpendicular to the electric field. This in turn changes the magnetic structure of the spin cycloid, making determination of the native anisotropy $\perp$ and $\parallel k$ impossible.\\

\begin{figure}[h!]%
    \centering
    \noindent\includegraphics[width=\textwidth]{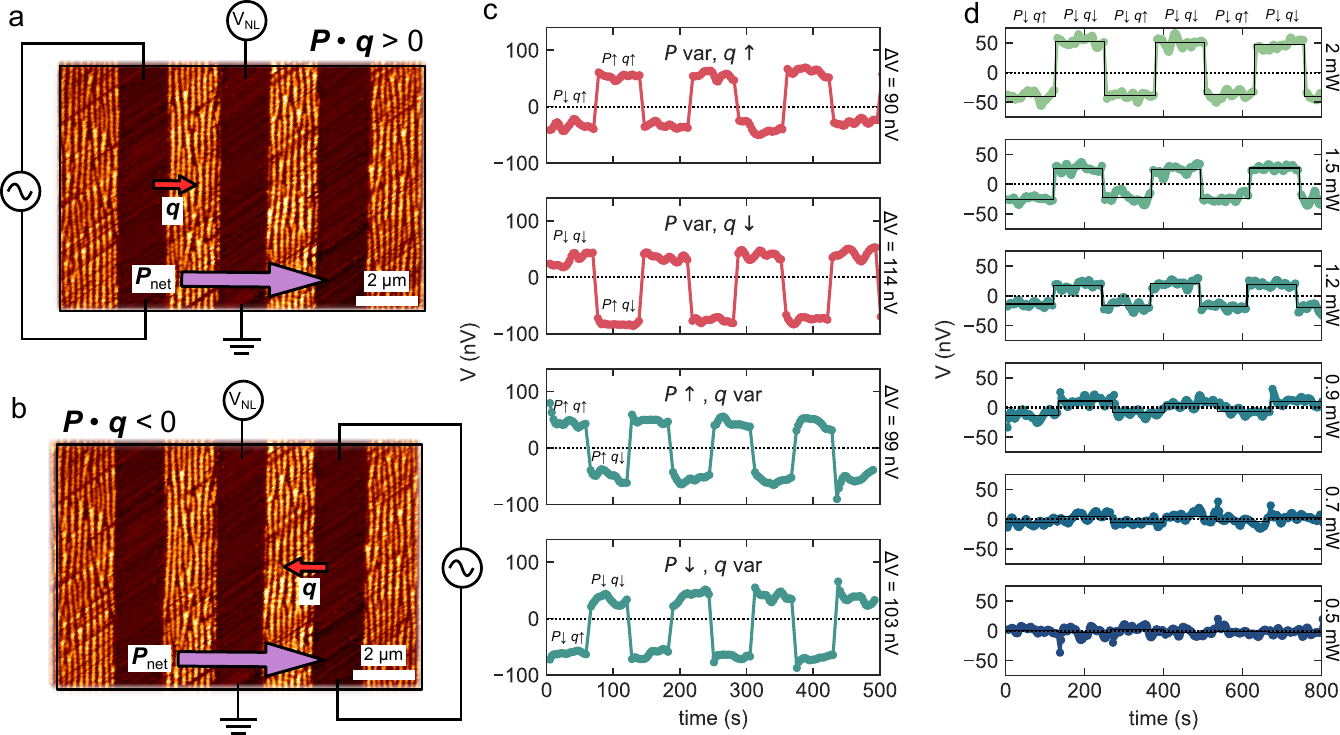}
    \caption{\textbf{Comparison of measurement schemes a,b} PFM and schematics of three terminal device for magnon-driven non-local voltage without an external electric field. The non-local voltage detector is fixed at the center and current is injected left or right to create two different $q$ while $P$ remains invariant. \textbf{c} Plots showing nonlocal signal in different configurations, switching $P$ under constant $q$ and vice versa. In all four permutations, $\Delta V \approx 100$ nV and the signs of the same states (e.g. $P\downarrow q\uparrow$) are consistent. This $\pm$ 10 nV signal may be due to the intrinsic device anisotropy, but should be mitigated when averaging across devices. \textbf{d} Power dependence of fixed $P$ variable $q$ measurements, showing the appropriate scaling and consistency of the sign.}
    \label{supfig:clvr}
\end{figure}

In samples where the electric field does not perturb the domain structure ($E \perp$ the domain walls), we find that the difference in $\Delta V_{NL}$ for the two polarization states is equivalent to the difference in $V_{NL}$ for the two configurations (\cref{supfig:clvr}b). This is reasonable following from the symmetry of the material. As there there is no net magnetization, and thus no net magnon polarization as in ferromagnetic magnon experiments, a difference in the nonlocal voltage is only expected to arise if the efficiency of magnon spin transport along $q$ is dependent on the underlying magnetic structure of the BFO. Using an electric field to switch the polarization direction, and thus the underlying magnetic structure, while leaving $q$ unchanged should be equivalent to switching $q$ while leaving $P$ and the underlying magnetic structure unchanged. In other words, we expect $\Delta V_{NL} = P\uparrow q\uparrow- P\downarrow q\uparrow \approx P\uparrow q\uparrow- P\uparrow q\downarrow$ when averaged across devices. This is shown in \cref{supfig:clvr}c. We note that while small differences in the heater and detector wire will lead to asymmetries in $V_{NL}$ unrelated to the magnon transport, averaging $\Delta V_{NL}$ across different devices negates this asymmetry.\\

\newpage
\section*{Supplementary Note 2: Micromagnetic modeling}

We employed MagneX, a substantial GPU-enabled micromagnetic simulation code package (available on \href{https://github.com/AMReX-Microelectronics/MagneX.git}{\sffamily \color{blue} \underline{GitHub}}), to phenomenologically explore the directionality of magnon transport in the cycloid. The time-domain Landau-Lifshitz-Gilbert equation is solved, incorporating exchange coupling, DMI coupling, anisotropy coupling, and demagnetization coupling.  A set of abstract ferromagnetic properties is utilized to phenomenologically represent the antiferromagnetic material.  \\

\begin{figure}[h!]
    \noindent\includegraphics[width=\textwidth]{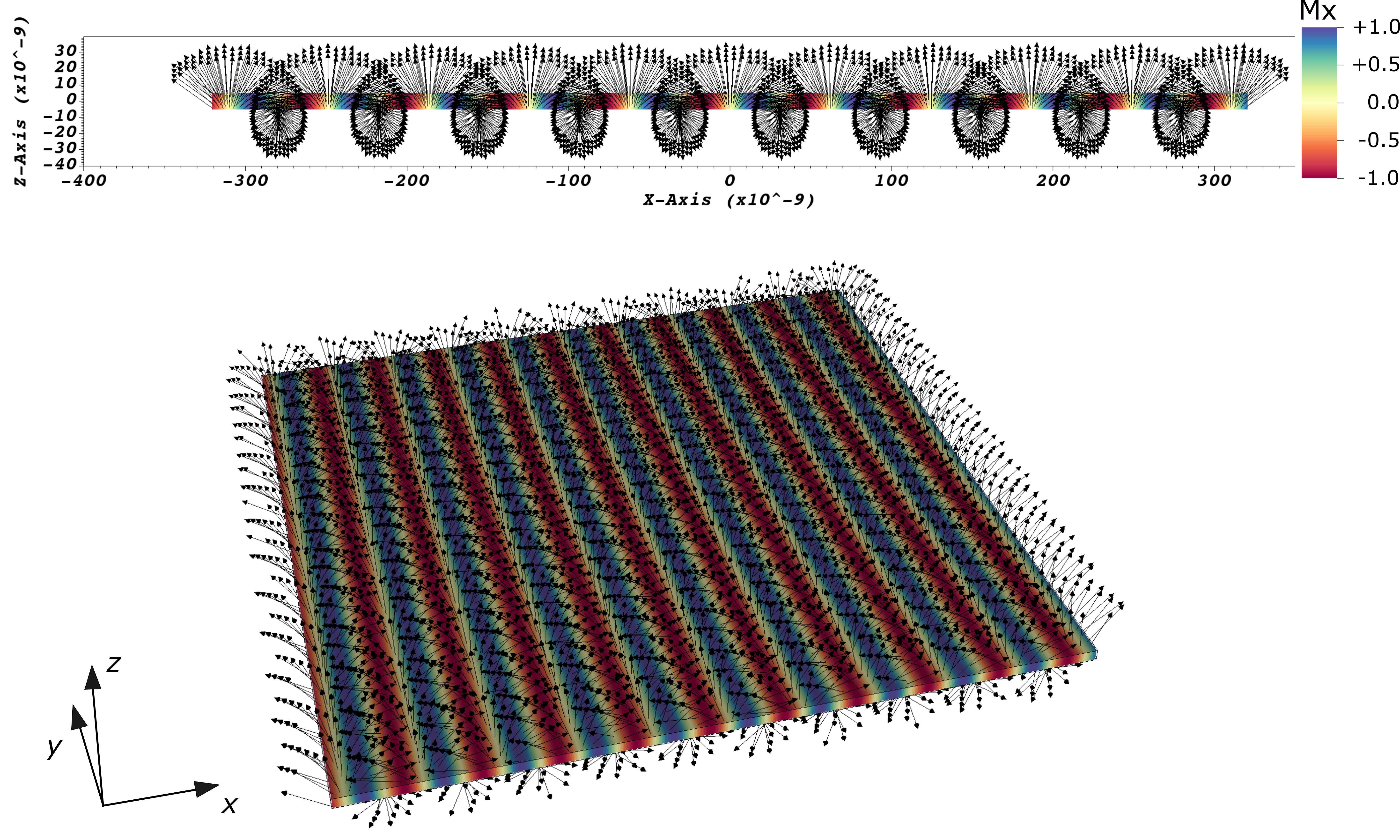}
    \caption{Initialized magnetization ground state in the simulation, with a period of 62 nm. Colored values show the magnitude of $M$ projected along $M_x$. Black arrows show the local $M$ vector. }
    \label{fig:modelM}
\end{figure}

The size of the sample is set to be 640 nm $\times$ 640 nm $\times$ 10 nm, with a 512$\times$512$\times$16 mesh resulting in grid sizes of $\Delta x=\Delta y=1.25$ nm, and $\Delta z = 0.625$ nm. The magnetization is pinned by a strong external applied magnetic field that mimics the cycloid texture, which gives the magnetization profile shown in \cref{fig:modelM}. The period of the cycloid is 62 nm. The Gilbert damping constant is set to be 0.0005, the exchange coupling constant is $3.0\times 10^{-12}$ J m$^{-1}$, and the DMI coupling constant is $1.0935\times 10^{-3}$ J m$^{-2}$. The anisotropic hard axis is set to the [111] direction, while the anisotropy constant is swept in the range of $1.0\times 10^{4}$ J m$^{-3}$  and $2.0\times 10^{6}$ J m$^{-3}$. The magnitude of the magnetization is normalized to one. \\

\begin{figure}[h!]
    \centering
    \noindent\includegraphics[width=0.7\textwidth]{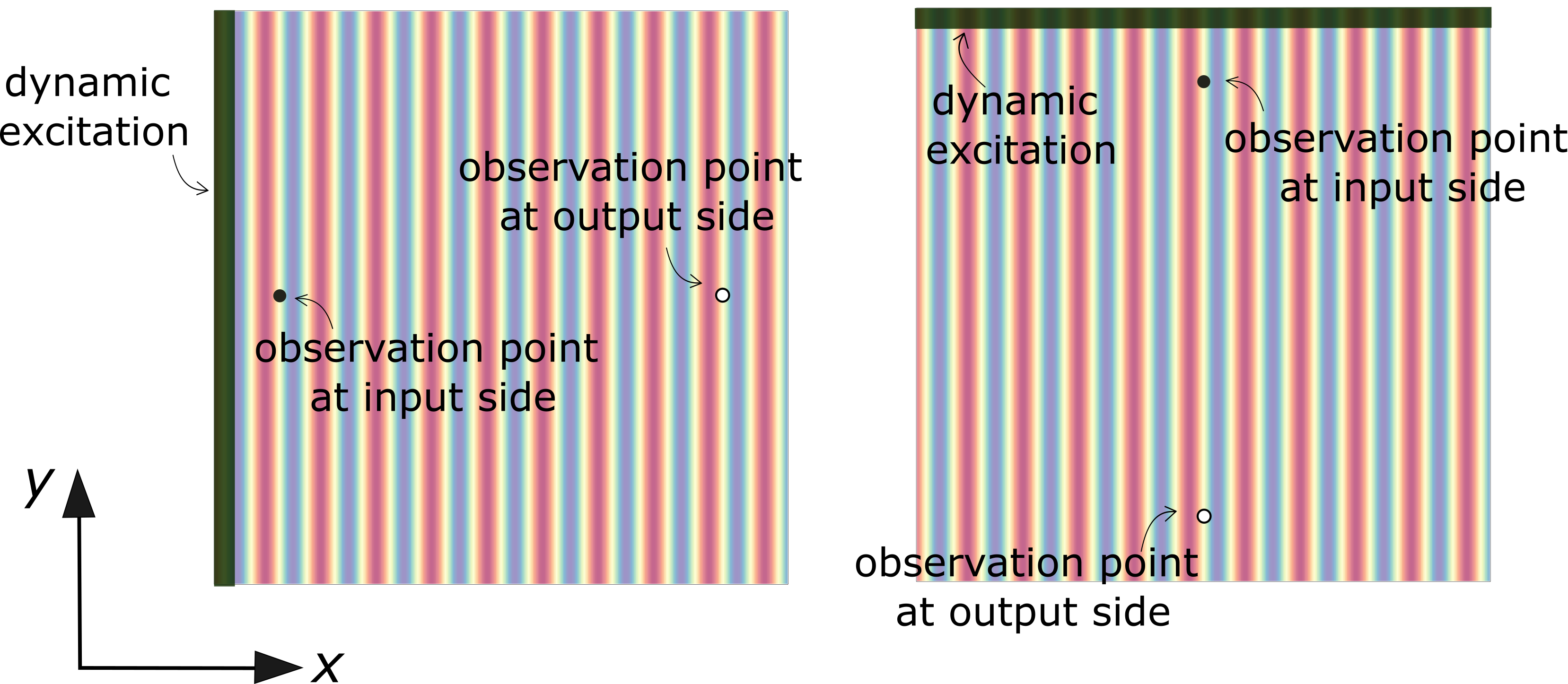}
    \caption{ Modeling setup. The left figure illustrates the measurement of magnon transport along the cycloid direction ($q \parallel k$), while the right figure depicts magnon transport along the uniform direction of the cycloid ($q \perp k$).}
    \label{fig:modelH}
\end{figure}

Starting from $t=0$, an additional time-dynamic external magnetic field is added on top of the existing pinning bias field. As shown in \cref{fig:modelH}, on one side of the sample, a modified Gaussian pulse in the effective magnetic field is applied in both the $x$ and $y$ direction, mimicking injected spins from the top SOC electrode. The pulse is in a mathematical form of $e^{-(t-t_0)^2/2T_0^2}\cos(2\pi f_0 t)$, where $f_0$ is 700 GHz. $T_0$ is chosen such that the excitation bandwidth is approximately 300 GHz. The output port is on the opposite side of the input port in the in-plane direction. By measuring the magnetization components, we can compute the spin wave transmission efficiency from the input port to the output port. \\

\captionsetup[figure]{format=plain,labelformat=default}

\begin{figure}[h!]
    \centering
    \noindent\includegraphics[width=1.0\textwidth]{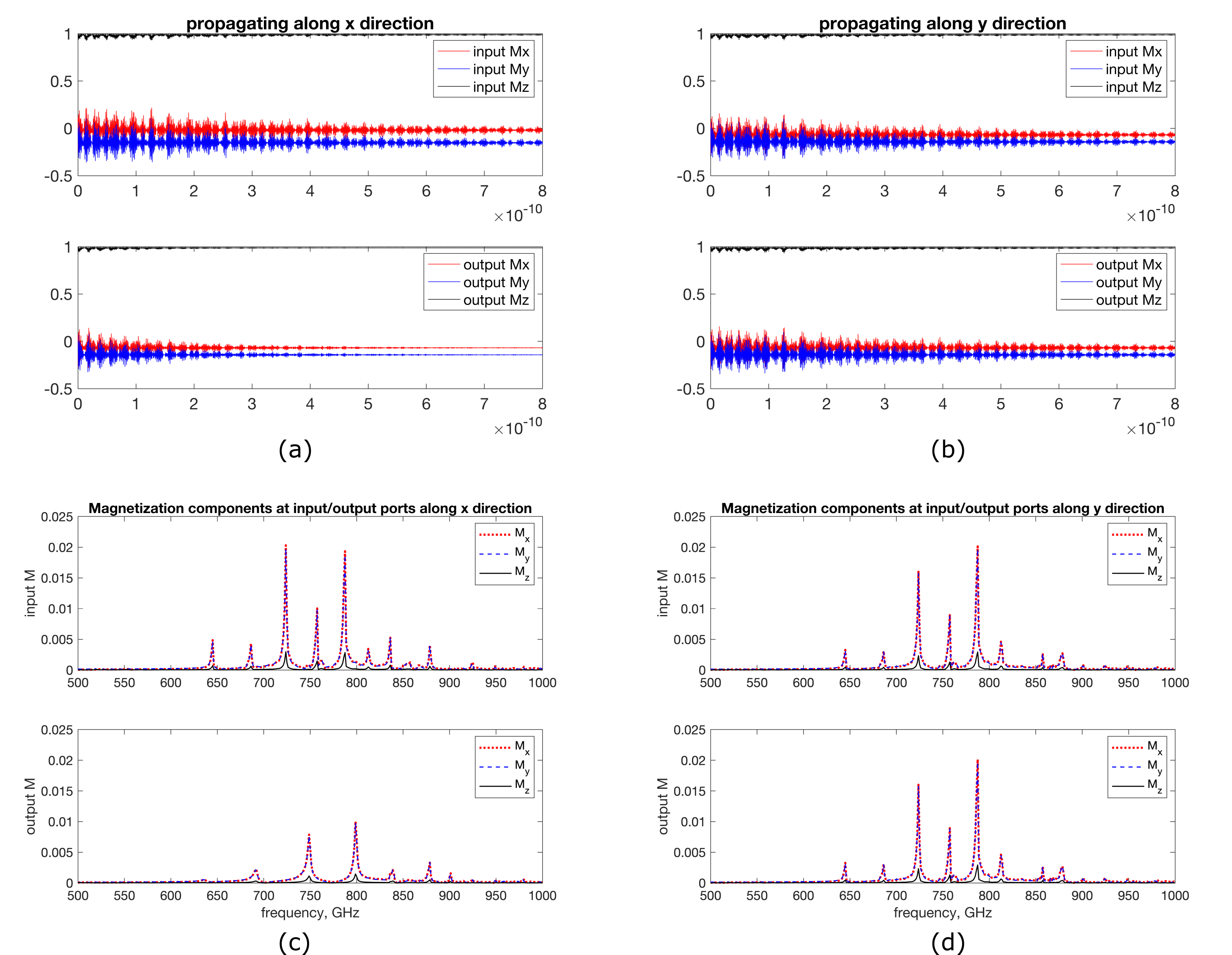}
    \caption{Time-domain and frequency-domain magnetization components at the input and output ports for both transport directions.}
    \label{fig:modelt}
\end{figure}

\cref{fig:modelt}a,b show the raw magnetization components at the input and output ports for both transport directions. The spots were selected where the ground-state magnetization is oriented in the out-of-plane direction, enabling straightforward analysis of the magnon excitation. These simulated time-varying magnetization components are then converted into the frequency domain through the Fourier transform, given in \cref{fig:modelt}c,d. \\

\captionsetup[figure]{format=plain,labelformat=default}

\begin{figure}[h!]
    \centering
    \noindent\includegraphics[width=0.7\textwidth]{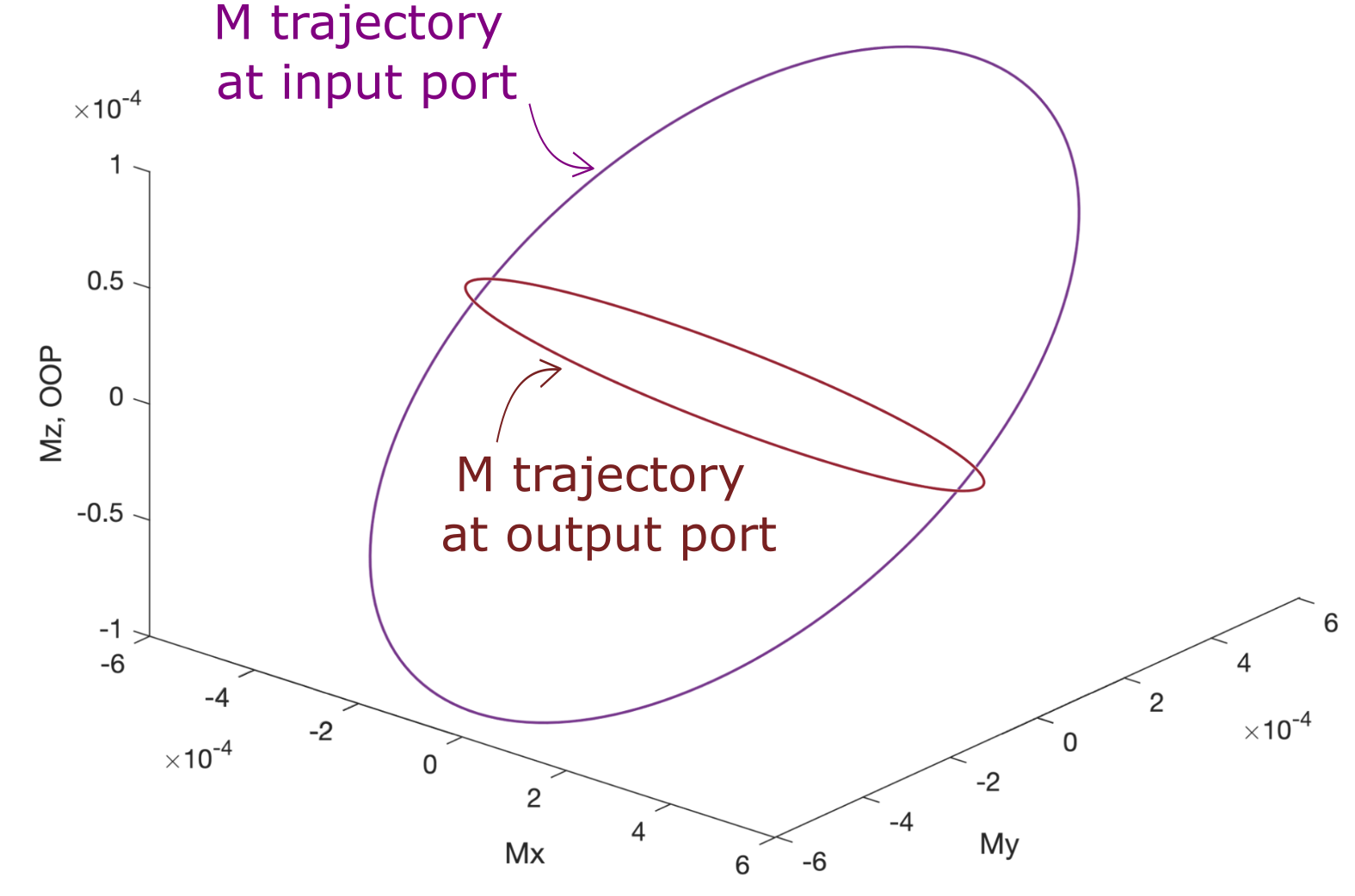}
    \caption{Magnetization trajectory at 700 GHz. Input and output ports are along x direction, i.e. the $K$ direction of the cycloid.}
    \label{fig:MTraj}
\end{figure}

To calculate the solid angles of the oscillating magnetization, we begin by expressing the spectral magnetization components at each frequency point in their time-harmonic form: $m = |M(f_0)|\times cos(2\pi f_0 t + \angle M(f_0))$.
This time-harmonic form provides an $M$ trajectory at each frequency point, with a cone area corresponding to the solid angle of the $M$ trajectory. For instance, \cref{fig:MTraj} illustrates the magnetization trajectory at 700 GHz for input and output ports along the cycloid $K$ direction. \\

\captionsetup[figure]{format=plain,labelformat=default}

Considering the small solid angle of magnetization, we simplify the calculation of the 3D cone area to a planar cross-sectional area $S(f)$ by integrating over all sectors forming the trajectory. The cone angles at each frequency are then determined by $\Omega(f) = S(f)/|M(f=0)|^2$, as illustrated in Extended Data Figure 3 in the main manuscript.
Subsequently, we calculate the magnon energy by integrating over its entire spectrum $E = \sum_{f} (\Omega(f)\times f)$.Finally, the transport efficiencies for both along-x and along-y scenarios are determined by computing the power efficiency $\xi = E_{\rm output} / E_{\rm input}$, leading to the plot in Figure 4c in the main manuscript. \\

\newpage
\section*{Supplementary Note 3: Analytical solution for the magnon bands}

Our theory for BiFeO$_{3}$ is based on the following continuum Hamiltonian \cite{park_structure_2014} and the corresponding Lagrangian \cite{baltz_antiferromagnetic_2018}, whose densities are respectively given by
\begin{align}
    \mathcal{H} &= A (\nabla \mathbf{n})^2 - \alpha\mathbf{P} \cdot [\mathbf{n}(\nabla \cdot\mathbf{n}) + \mathbf{n}\times(\nabla \times \mathbf{n})] - 2\beta M_0 \mathbf{P}\cdot(\mathbf{m}\times\mathbf{n}) - K_u n_c^2 + \lambda \mathbf{m}^2 \, , \\
    \mathcal{L} &= s\mathbf{m}\cdot(\mathbf{n}\times \dot{\mathbf{n}}) - \mathcal{H} \, .
\end{align}
Here, the magnetic state is described by the Neel order parameter $\mathbf{n}(\mathbf{r}, t)$ whose length is unity and the net magnetization $\mathbf{m}(\mathbf{r}, t)$ which is perpendicular to the Neel order parameter, i.e., $\mathbf{n} \cdot \mathbf{m} = 0$. The coefficients in the continuum Hamiltonian are related to the parameters in the microscopic spin Hamiltonian as follows: $\lambda = \frac{6S^2}{V}(-3\mathcal{J}+6\mathcal{J'}) \, , A = \frac{6S^2} {V}a_{hex}^2\frac{3\mathcal{J}-4\mathcal{J'}}{4} \, , \alpha = \frac{6S^2}{V} \frac{1}{P_s}a_{hex}^2\mathcal{D}_u \, , \beta = \frac{6S^2}{V} \frac{1}{M_0P_s}\mathcal{D}_c$, and $K_u =  \frac{6S^2}{V} \mathcal{K}$, where $\mathcal{J} = 4.38$ meV, $\mathcal{J}' = 0.15$ meV, $\mathcal{D}_u = 0.11$ meV, $\mathcal{D}_c = 0.05$ meV, $\mathcal{K} = 0.003$ meV, $S=\sqrt{\frac{5}{2}(\frac{5}{2}+1)}$, $a_{hex} = 5.58\times10^{-8}$ cm, and $V = 375.9 \times 10^{-24}$ cm$^{3}$ \cite{park_structure_2014}. The microscopic spin Hamiltonian can be found in Ref. \cite{park_structure_2014}. It is convenient to use the spherical angles $\theta(\mathbf{r}, t)$ and $\phi(\mathbf{r}, t)$ to represent $\mathbf{n} = (\sin\theta\cos\phi,\sin\theta\sin\phi,\cos\theta)$ and $m_\theta(\mathbf{r}, t)$ and $m_\phi(\mathbf{r}, t)$ to represent $\mathbf{m}= (m_\theta\cos\theta\cos\phi-m_\phi\sin\phi, m_\phi\cos\phi+ m_\theta\cos\theta\sin\phi, -m_\theta\sin\theta)$. In terms of these variables, the Euler-Lagrange equation of the Lagrangian is given by 
\begin{align}
    -sm_\theta \partial_t \phi \cos\theta -s\partial_t m_\phi+ 2A(\nabla^2\theta)+2\alpha P_z \sin^2\theta\left(\sin\phi(\partial_x\phi)-\cos\phi(\partial_y\phi)\right) \quad & \nonumber\\
    - \sin2\theta (A(\nabla\phi)^2 + K_u) - 2\beta M_0 P_z m_\phi \cos\theta &= 0 \, , \\
    s\partial_t(m_\theta \sin\theta) + 2A\nabla\cdot(\sin^2 \theta(\nabla\phi)) - 2\alpha P_z \sin^2 \theta(\sin\phi(\partial_x\theta) -\cos\phi(\partial_y\theta)) &= 0 \, , \\
    -s \partial_t \phi \sin\theta - \frac{m_\theta}{\chi} &= 0 \, , \\
    s \partial_t \theta - \frac{m_\phi}{\chi} - 2\beta M_0 P_z \sin \theta &= 0 \, .
\end{align}\\

The ground-state configuration $\mathbf{n}_0 (\mathbf{r})$ and $\mathbf{m}_0 (\mathbf{r})$ can be obtained by solving the Euler-Lagrange equation with no time dependence. The resulting ground state of BiFeO$_{3}$ has an anharmonic spin cycloid structure as well known, e.g., in Refs. \cite{sosnowska_origin_1995, kadomtseva_phase_2006, park_structure_2014}. Here, we focus on the cases where $\beta = 0$, as often done due to its smallness \cite{park_structure_2014}. The explicit expression for a ground state is given by
\begin{align}
    \phi = \text{const.} , \quad \theta = \mathrm{am} \left( \sqrt{\frac{C}{A}}x,\frac{K_u}{C}\right) - \frac{\pi}{2} \, ,
\end{align}
with vanishing $\mathbf{m}$, where $\mathrm{am}(u, m)$ is the Jacobi amplitude function and $C$ is the integral constant determined by minimizing the free energy of the aforementioned solution with respect to $C$~\cite{park_structure_2014}. Here, the spin-cycloid direction is chosen to be the $x$-axis without loss of generality.\\

By solving the time-dependent Euler-Lagrange equation for the small fluctuations, $\delta \mathbf{n} = \mathbf{n} - \mathbf{n}_0$ and $\delta \mathbf{m} = \mathbf{m} - \mathbf{m}_0$, on top of the ground state up to linear order in the fluctuations $\delta \mathbf{n}$ and $\delta \mathbf{m}$, we can obtain two spin-wave modes: in-plane mode with spin fluctuations within the spin-cycloid plane and out-of-plane mode whose spin fluctuations are perpendicular to the spin-cycloid plane. The corresponding equations of motion for $\delta \theta, \delta \phi, \delta  m_\theta$, and $\delta m_\phi$ are given by 
\begin{align}
    -s\partial_t \delta m_\phi + 2A(\nabla^2\delta\theta) +2\alpha P_z (\sin^2\theta_g)\left(\sin\phi_g(\partial_x \delta\phi)-\cos\phi_g(\partial_y \delta\phi)\right) & \nonumber\\- (2\delta\theta\cos2\theta_g) (K_u) = 0 \, , & \\
    s\partial_t(\delta m_\theta \sin\theta_g) + 2A\nabla\cdot(\sin^2\theta_g(\nabla(\delta \phi))) & \nonumber\\
    - 2\alpha P_z (\sin^2\theta_g)(\sin\phi_g(\partial_x \delta\theta) + \delta\phi\cos\phi_g (\partial_x\theta_g) -\cos\phi_g (\partial_y \delta\theta) + \delta\phi\sin\phi_g (\partial_y\theta_g)) = 0 \, , &  \\
    -s\delta\Dot{\phi}\sin\theta_g - \frac{\delta m_\theta}{\chi} = 0 \, , & \\
    \delta\Dot{\theta} - \frac{\delta m_\phi}{s\chi} = 0 \, . &
\end{align}\\

The translational symmetry along the $z$ axis allows us to use the plane-wave ansatz $\delta \theta (x, z, t) = \delta \theta (x; k_z, \omega) \exp(i k_z z - i \omega t)$ and $\delta \eta (x, z, t) = \delta \eta (x; k_z, \omega) \exp(i k_z z - i \omega t)$ with $\delta \eta = \sin \theta_g \delta \phi$, in terms of which the above four equations can be combined into the following two equations for $\delta \theta(x; k_z, \omega)$ and $\delta \eta (x; k_z, \omega)$:
\begin{align}
    [(s^2\omega^2 \chi - 2Ak_z^2) + 2A\partial_x^2 - 2 K_u \cos2\theta_g]\delta\theta &= 0 \, , \\
     \left[(s^2\omega^2 \chi - 2 A k_z^2) + 2 A \partial_x^2 - 4 K_u \cos^2 (\theta_g) + 2 C - 2\alpha P_z (\partial_x\theta_g)\right] \delta \eta &= 0 \, ,
\end{align}
where $\delta\eta = \sin\theta_g\delta\phi$. By solving these equations numerically, we obtained the band structures of in-plane magnons and out-of-plane magnons as shown in the Figure 3 of the main manuscript. It is apparent that the band structure for the two modes is anisotropic. We remark that the anisotropy is the most visible in the lowest bands, which makes sense since low-energy magnons are expected to be more affected by the spin-cycloid structure than high-energy magnons and therefore exhibit anisotropic behavior more strongly.

\newpage
\section*{references}